%% file: main.tex
\documentclass[conference]{IEEEtran}
\IEEEoverridecommandlockouts
\usepackage{authblk}
\input{imports.tex}

\begin{document}

\title{Comprehensive Benchmarking of Binary Neural Networks on NVM Crossbar Architectures}

\author[1]{Ruirong Huang}
\author[1]{Zichao Yue}
\author[1]{Caroline Huang}
\author[2]{Janarbek Matai}
\author[1]{Zhiru Zhang}
\affil[1]{Department of Electrical and Computer Engineering, Cornell University}
\affil[2]{Qualcomm Technologies, Inc.}

\maketitle
\input{Sections/abstract}

\input{Sections/introduction}
\input{Sections/background}
\input{Sections/design}
\input{Sections/experiments}

\input{Sections/conclusion}

\input{Sections/acknowledgement}
    
\bibliographystyle{IEEEtran}
\bibliography{macro.bib, all.bib}

\end{document}

%% file: imports.tex
\usepackage{cite}
\usepackage{url}
\usepackage{amsmath,amssymb,amsfonts}
\usepackage{algorithmic}
\usepackage{graphicx}
\usepackage{textcomp}
\usepackage{xcolor}
\usepackage[normalem]{ulem}
\usepackage[utf8]{inputenc}
\usepackage{subcaption}
\usepackage{diagbox}

\usepackage{multirow}
\usepackage{booktabs}

%% file: Sections/abstract.tex
\begin{abstract}
Non-volatile memory (NVM) crossbars have been identified as a promising technology, for accelerating important machine learning operations, with matrix-vector multiplication being a key example. Binary neural networks (BNNs) are especially well-suited for use with NVM crossbars due to their use of a low-bitwidth representation for both activations and weights. However, the aggressive quantization of BNNs can result in suboptimal accuracy, and the analog effects of NVM crossbars can further degrade the accuracy during inference.

This paper presents a comprehensive study that benchmarks BNNs trained and validated on ImageNet and deployed on NeuroSim, a simulator for NVM-crossbar-based PIM architecture. Our study analyzes the impact of various parameters, such as input precision and ADC resolution, on both the accuracy of the inference and the hardware performance metrics. We have found that an ADC resolution of 8-bit with an input precision of 4-bit achieves near-optimal accuracy compared to the original BNNs. In addition, we have identified bottleneck components in the PIM architecture that affect area, latency, and energy consumption, and we demonstrate the impact that different BNN layers have on hardware performance.

\end{abstract}


%% file: Sections/introduction.tex

\section{Introduction}
\label{sec:intro}

Non-volatile memory (NVM) crossbars have become popular as a type of analog computing substrate used in processing-in-memory (PIM) architectures \cite{shafiee_isaac_2016, ankit_puma_2019, ankit2020panther, boybat_neuromorphic_2018}. 
These crossbars consist of an array of NVM cells and perform matrix-vector multiplication (MxV) in the analog domain using Ohm's law and Kirchhoff's current law. By processing data in-situ, the need for data movement is reduced, resulting in increased performance and energy efficiency. This approach is gaining traction in hardware acceleration for machine learning, where MxV is a critical computational kernel. 

However, supporting high-precision neural networks in NVM crossbars can be challenging because of the need for digital-to-analog converters (DAC) and analog-to-digital converters (ADC) to convert signals from the digital domain to the analog domain for processing and vice versa. High-precision neural networks require high-resolution DAC and ADC, which consume more energy and can offset the benefits of in-situ computing. As a result, researchers often resort to quantization techniques to reduce precision while maintaining acceptable accuracy levels, making quantized neural networks a more viable option for NVM crossbars \cite{hung2021challenges,yu2018neuro,chakraborty2020resistive}.

Binary neural networks (BNNs) are a type of quantized neural network that uses a single bit to represent weights and activations~\cite{ zhang_fracbnn_2020, hubara_binarized_nodate, li_bcnn_2021, qin_binary_2020, liang_fp-bnn_2018, lin_towards_nodate}. 
The extreme quantization of BNNs can lower both computational complexity and storage requirements. When deployed in NVM crossbars, BNNs require only a simple inverter for input activation, and the shift-and-add circuit commonly used for weight bit-slicing can be eliminated, leading to greater energy efficiency. This synergy between BNNs and NVM crossbars creates a promising scenario for deploying many emerging applications at the edge settings with strict resource and latency constraints. The synergy between BNNs and NVM crossbars creates a promising scenario for deploying a variety of emerging ML workloads at the edge with strict resource and latency constraints.

While deploying BNNs in NVM crossbars holds potential benefits, it is crucial to pay close attention to the possible negative effects. BNNs typically have lower inference accuracy due to their aggressive quantization nature compared to high-precision counterparts. The analog effects of NVM crossbars, such as insufficient ADC resolution and memory non-idealities, can further decrease the inference accuracy. Additionally, the configuration of BNNs, including model structure and input precision, can significantly impact the hardware performance of NVM crossbars. Improper choices of the model configuration can offset the advantages of energy efficiency gained by using the NVM crossbar.

Several simulators have been proposed for NVM-crossbar-based neural network accelerators \cite{crosssim, He2019NIA, jain2020rxnn, roy2021txsim, zhu2020mnsim, peng2019dnn+} to facilitate fast design space exploration and performance evaluation. These simulators typically encompass different levels of hardware modeling, including memory, circuit, and chip, with varying degrees of fidelity. For instance, NeuroSim \cite{peng2019dnn+} provides detailed circuit-level modeling, while MNSIM \cite{zhu2020mnsim} offers more flexible chip-level exploration options. Previous studies have utilized these simulators to explore the potential of NVM crossbars for BNN inference, primarily focusing on simpler BNNs \cite{sun_computing--memory_2018, liang_fp-bnn_2018, zahedi_bcim_2022} and small datasets such as MNIST \cite{sun_fully_2018, zahedi_bcim_2022} and CIFAR-10 \cite{sun_xnor-rram_2018, yin_xnor-sram_2020, liang_fp-bnn_2018, kim_area-efficient_2019}. However, a comprehensive benchmarking study is still lacking to investigate the effectiveness of NVM crossbars on more complex BNN models using realistic datasets. Such a benchmark could provide valuable insights into the performance bottleneck of NVM crossbar hardware components and which BNN features are best suited for use in NVM crossbars. 

In this paper, we use the open-source simulator NeuroSim \cite{peng2019dnn+} to conduct a comprehensive benchmark of modern BNNs implemented on NVM-crossbar-based PIM architecture. The contributions of this paper are as follows.

\begin{itemize}
    \item This study provides the first comprehensive benchmarking of three realistic BNN models (BAlexNet, BResNet, and BDenseNet) on an NVM-crossbar-based PIM architecture, trained and validated on the ImageNet dataset. We analyze the impact of various parameters, including input precision and ADC resolution, on both inference accuracy and chip performance, such as chip area, latency, energy consumption, and throughput.

    \item We analyze the relationship between inference accuracy, input precision, and ADC resolution, and demonstrate that increasing the ADC resolution and input precision can generally result in higher accuracy but lower throughput. Our study shows that an ADC resolution of 8-bit with an input precision of 4-bit offers a negligible accuracy loss compared to the same BNN model evaluated on a GPU with float32 first-layer activation, while achieving a better energy efficiency trade-off.

    \item 
    We identify the hardware components that bottleneck the performance of the NVM crossbar architecture. Specifically, our work demonstrates that the area of the ADC grows exponentially with increasing resolution, becoming the dominant part of the chip in terms of area. Additionally, we find that buffer latency and interconnection dynamic energy are major contributors to latency and energy consumption, respectively.

    \item We demonstrate how different model structures impact inference accuracy and hardware performance. Our findings reveal that for deeper models, the pipelined architecture of the NVM accelerator
    can significantly reduce latency, but large fully-connected layers become the bottleneck of chip performance, especially in terms of latency.
\end{itemize}



%% file: Sections/background.tex
\section{Background}
\label{sec:background}

\subsection{Binary Neural Network}
Binary Neural Network \cite{zhang_fracbnn_2020, hubara_binarized_nodate, li_bcnn_2021, qin_binary_2020, liang_fp-bnn_2018, lin_towards_nodate} is a type of neural network with binary weights and activations. In a BNN, each weight and activation value is either +1 or -1, which  takes only 1 bit of memory, rather than a floating point or integer value, which takes 32 or 16 bits.

The use of binary weights and activations in BNNs offers several advantages over traditional neural networks, including reduced memory requirements, faster processing speed, and improved energy efficiency. Additionally, BNNs have been shown to be more robust to noisy inputs and adversarial attacks compared to traditional neural networks. This is due to the reduced sensitivity of binary values to small perturbations in the input.

However, BNNs also have some limitations compared to traditional neural networks. The binary weights and activations result in a loss of expressiveness and accuracy compared to their higher bit-precision counterparts. Additionally, training BNNs can be challenging, as the non-differentiable binary activation function can make it difficult to perform back-propagation and update the weights.



\subsection{Processing-in-Memory}
Processing-in-Memory (PIM) architecture is a computer architecture that endows the memory module with computing capability \cite{schuman_survey_2017, aly_modern_2023, caminal_cape_2021}. One of the main benefits of PIM is reduced data movement and communication overhead. In traditional architectures, data must be transferred from memory to the processor for processing, which significantly increases the energy consumption and latency. PIM eliminates this transfer, as the data is processed directly in the memory, hence improving performance and energy efficiency. In this paper, we focus on an emerging PIM device known as the NVM crossbar.

\subsubsection{Embedded Non-volatile Memory}
Non-volatile memory (NVM) is a type of memory that can retain data even when power is not supplied. Several emerging NVM technologies, such as resistive random access memory (RRAM), phase change random access memory (PCM), and magnetoresistive random access memory (MRAM), offer comparable or shorter latency compared to dynamic random access memory (DRAM), along with higher density than DRAM and static random access memory (SRAM) \cite{boukhobza2017emerging}. Embedded non-volatile memory (eNVM) specifically refers to NVM integrated into a system-on-chip (SoC) or an application-specific integrated circuit (ASIC) \cite{ielmini_-memory_2018, sebastian_memory_2020, burr_neuromorphic_2017}. In addition to the advantages of NVM, eNVM can be manufactured using standard CMOS processes, making it relatively low-cost and easy to integrate into existing designs. These benefits make eNVM-based crossbar an attractive option for ML workloads. In this paper, we focus on RRAM, which uses material that can be switched between two or multiple resistance states to store one or more bits of information \cite{roy_towards_2019, boybat_neuromorphic_2018, burr_neuromorphic_2017}. We use the 22nm RRAM model provided in NeuroSim for our benchmark.

\subsubsection{NVM Crossbar}
NVM crossbar \cite{ankit2020panther,ankit_puma_2019,shafiee_isaac_2016,boybat_neuromorphic_2018} is a type of PIM architecture that enables parallel processing and data storage by integrating memory and computation functions within a single device. In the crossbar architecture, the memory cells are arranged in a two-dimensional array of rows and columns, and each cell is connected to both a  \emph{wordline} and a  \emph{bitline}. One of the main advantages of a crossbar architecture is its ability to perform massively parallel operations, which can greatly accelerate computation and reduce power consumption. In addition, the crossbar can be implemented using a variety of NVM technologies, such as RRAM, PCM or MRAM, which can provide both high-density and low-power storage.

Matrix-vector multiplication is usually the dominant computation in ML workloads, which can be done in parallel through crossbar devices \cite{ankit_puma_2019, chi_prime_2016, zahedi_bcim_2022} with each crosspoint storing one weight. With the input activations applied as voltage to horizontal lines, the vector-matrix multiplication is automatically down through the principles of Ohm's law and Kirchhoff's current law. By sensing the amplitude of the output current at each vertical line, we can read out the vector-matrix multiplication results in parallel.

\subsubsection{NeuroSim}
NeuroSim is a simulation platform designed for modeling and analyzing the performance of deep neural network (DNN) accelerators adopting NVM-crossbar-based architectures \cite{sun_xnor-rram_2018, sun_fully_2018, sun_computing--memory_2018}. NeuroSim provides hardware models that cover different levels of abstraction, including memory, circuit, and chip architecture. Compared to other open-source simulators such as CrossSim \cite{crosssim}, PytorX \cite{He2019NIA} and MNSIM \cite{zhu2020mnsim}, NeuroSim offers more detailed circuit-level models, allowing for accurate estimation of inference accuracy and hardware performance, including area, latency, and energy consumption. 
However, it is important to note that one limitation of NeuroSim is that the accelerator floorplans generated for different neural networks are model-specific. 



%% file: Sections/design.tex
\section{Benchmark Design}
\label{sec:design}   

In this section, we present our benchmark design, which involves three BNN models: BAlexnet, BResnet18, BDensenet28, each with two focuses: inference accuracy and hardware performance. First, we introduce the models used in our experiments, and then we detail our benchmark design and methodology.

\subsection{Models}
NeuroSim provides a PyTorch wrapper for simulating the inference accuracy of different neural network models. We build BAlexNet, BResNet18, and BDenseNet28 in PyTorch referring to their TensorFlow-version from Larq Zoo \cite{geiger_larq_2020}. 

\subsubsection{AlexNet}
AlexNet \cite{krizhevsky_imagenet_2017} is a convolutional neural network (CNN) architecture consisting of five convolution layers and three large fully-connected layers. The convolution layers employ a combination of filters of varying sizes and strides, which are then followed by max-pooling layers. The fully-connected layers are succeeded by a softmax layer that produces the class probabilities.

\subsubsection{ResNet}
ResNet, \cite{he_deep_2016} short for "Residual Network", is a CNN architecture that introduces residual connections, which allow information to be passed directly from one layer to another. This is done by adding the output of a previous layer to the input of a subsequent layer, allowing the network to learn with a much deeper size and produce better accuracy. Various versions of ResNet exist with differing numbers of layers. In our benchmark, we employ ResNet18.

\subsubsection{DenseNet}
DenseNet, \cite{huang_densely_2017} short for "Densely Connected CNN", is a deep neural network architecture that introduces densely connected blocks, which are made up of several convolution layers with a fixed number of filters. Each layer receives the feature maps of all preceding layers as inputs, concatenated together along the depth dimension. This dense connectivity pattern allows the network to reuse features learned by earlier layers, which results in better parameter efficiency and gradient flow. In our benchmark, we utilize DenseNet28.

\subsection{Inference Accuracy}
Accuracy is one of the biggest concerns in BNNs, while model inference is the most common use case for edge machine learning devices. In order to get realistic results of inference accuracy for BNN models on large datasets, we first train all three models on ImageNet using binary weights for all layers and binary activations for all but the first layer. Next, we evaluate the trained model using the NeuroSim simulator by replacing key functions such as nn.conv2d with APIs provided by NeuroSim that account for hardware effects. However, since NeuroSim is not specifically designed for BNNs, it employs a high-precision floating-point input for the first layer, with bit-serialization modules used to process activations in the subsequent layers. To better simulate BNNs, we introduce a bit-serialization module to process the input of the first layer and eliminate this module from all subsequent layers, given that the activations of these layers are already binarized. We also employ dynamic quantization of input data. To further investigate, we focus on two parameters with the most significant impact on accuracy: first-layer input precision and the resolution of ADCs.

\subsection{Hardware Performance}
To quantify the performance of BNNs deployed in NVM crossbar architectures, we first provide layer-level model structures for NeuroSim, since it's a model-specific simulator. Each layer includes the dimension of its input activation and weight, as well as parameters for any potential subsequent pooling and activation module (i.e. relu). Furthermore, we collect the intermediate activation and weight data for each layer from the trained model, as this information is crucial for NeuroSim to generate more precise estimates based on the distribution of each layer's input activation and weight. NeuroSim offers different modes catering to distinct design considerations. In our benchmark, we select the \textit{XnorParallel} mode, which employs two memory cells to represent a weight of either +1 or -1, making it suitable for BNNs. Moreover, we activate the pipeline mode in NeuroSim to leverage the pipelined architecture of the chip, thereby achieving enhanced throughput. To demonstrate the impact of ADC resolution on both accuracy and hardware performance, we perform tests on each model using various ADC resolutions and collect their hardware performance data, including chip area, latency, energy consumption, throughput, and efficiency.


%% file: Sections/experiments.tex
\section{Experiments}
\label{sec:experiments}

This section provides a detailed explanation of our experiment parameters and results, divided into four parts: inference accuracy, area, latency, and energy, each with four subsections corresponding to the three BNN models. Additionally, there is a Put-It-All-Together section that compares the results of the three models. Finally, we analyze the throughput and efficiency of the three models.

\subsection{Inference Accuracy}

\subsubsection{AlexNet}

\begin{table}[htbp]
    \caption{\textbf{Inference accuracy for Binary AlexNet with different input precisions and ADC resolutions} --- The inference accuracy on GPU with float32 first-layer activation is 36.11\%.}
    \label{tab:acc_alexnet}
    \centering
    \resizebox{0.4\textwidth}{!}{
        \input{Assets/acc/alexnet.tex}
    }
\end{table}

\begin{figure}[htbp]
  \centering
   \includegraphics[width=\linewidth]
  {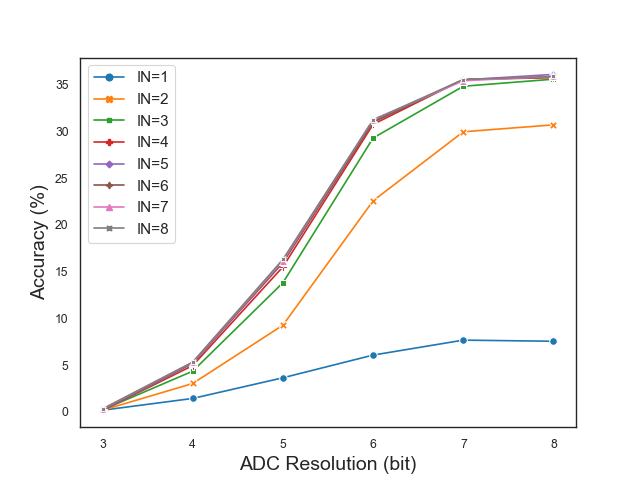}
  \caption{\textbf{Binary AlexNet inference accuracy vs. ADC resolution} --- Each line corresponds to a different first-layer input precision.}
  \label{fig:acc_alexnet_adc}
\end{figure}

\begin{figure}[htbp]
  \centering
  \includegraphics[width=\linewidth]{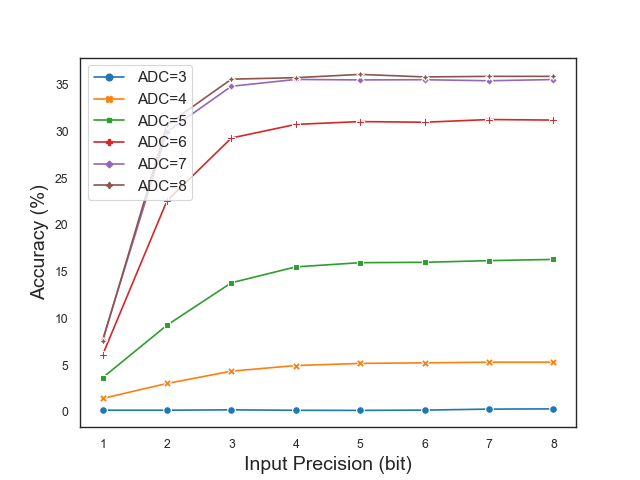}
  
  \caption{\textbf{Binary AlexNet inference accuracy vs. first layer input precision} --- Each line corresponds to a different ADC resolution.}
  \label{fig:acc_alexnet_wlinput}
\end{figure}

Table \ref{tab:acc_alexnet} presents the inference accuracy for Binary AlexNet across different ADC resolutions (3-8) and first-layer input precisions (1-8). The same data is used to generate Figure \ref{fig:acc_alexnet_adc} and Figure \ref{fig:acc_alexnet_wlinput}, which visually depict the trend of accuracy change. In Figure \ref{fig:acc_alexnet_adc}, the accuracy for input precision of 1 increases with ADC resolution but remains low at the highest precision (7.55\% for ADC resolution=8). Accuracy improves significantly with an input precision of 2, reaching a maximum of 30.71\%. With an input precision of 3, accuracy reaches an asymptote, and increasing input precision beyond 3 has little effect on accuracy.

Figure \ref{fig:acc_alexnet_wlinput} also shows that accuracy monotonically increases with ADC resolution, although this trend stops when ADC resolution reaches 7. Overall, input precision of 4-bit with ADC resolution of 8-bit provides an acceptable accuracy of 35.60\%, which is only 0.8\% lower than the maximum accuracy of 35.9\% when both parameters are set to 8, and is only 1.4\% lower than the accuracy of 36.11\% achieved on GPU with float32 first-layer precision.

\subsubsection{ResNet} 

\begin{table}[htbp]
    \caption{\textbf{Inference accuracy for Binary ResNet  with different input precisions and ADC resolutions} --- The inference accuracy on GPU with float32 first-layer activation is 54.02\%.}
    \label{tab:acc_resnet}
    \centering
    \resizebox{0.4\textwidth}{!}{
        \input{Assets/acc/resnet.tex}
    }
\end{table}

\begin{figure}[htbp]
  \centering
   \includegraphics[width=\linewidth]
  {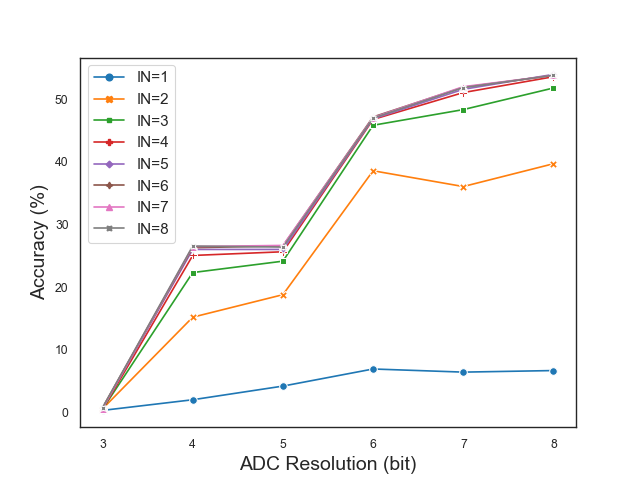}
  \caption{\textbf{Binary ResNet inference accuracy vs. ADC resolution} --- Each line corresponds to a different first-layer input precision.}
  \label{fig:acc_resnet_adc}
\end{figure}

\begin{figure}[htbp]
  \centering
  \includegraphics[width=\linewidth]{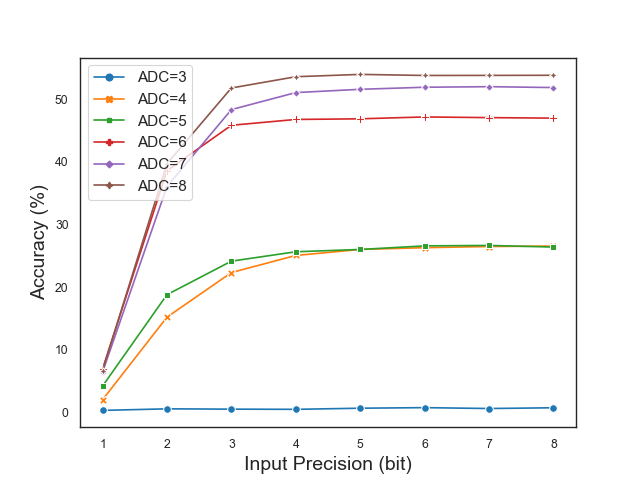}
  \caption{\textbf{Binary ResNet inference accuracy vs. first layer input precision} --- Each line corresponds to a different ADC resolution.}
  \label{fig:acc_resnet_wlinput}
\end{figure}

Table \ref{tab:acc_resnet} lists the inference accuracy according to different ADC resolutions (3-8) and first-layer input precisions (1-8) for Binary ResNet. The same data is used to draw Figure \ref{fig:acc_resnet_adc} and Figure \ref{fig:acc_resnet_wlinput}. Similar to the results of AlexNet, in Figure \ref{fig:acc_resnet_adc}, the accuracy is still quite low when input precision is 1. It gets a great improvement when input precision increases to 2, where the max accuracy reaches 39.72\%. 




Overall, input precision of 4-bit with ADC resolution of 8-bit provides an acceptable accuracy of 53.61\%, which is only 0.4\% lower than the maximum accuracy of 53.84\% when both parameters are set to 8, and is only 0.7\% lower than the accuracy of 54.02\% achieved on GPU with float32 first-layer precision.

\subsubsection{DenseNet}

\begin{table}[htbp] 
    \caption{\textbf{Inference accuracy for Binary DenseNet with different input precisions and ADC resolutions} --- The inference accuracy on GPU with float32 first-layer activation is 56.15\%.}
    \label{tab:acc_densenet}
    \centering
    \resizebox{0.4\textwidth}{!}{
        \input{Assets/acc/densenet.tex}
    }
\end{table}

Table \ref{tab:acc_densenet} presents the inference accuracy for Binary DenseNet across different ADC resolutions (3-8) and first-layer input precisions (1-8). Similar to the result of ResNet, the accuracy when input precision is 1 is still quite low. It gets great improvement when input precision increases to 2, where the max accuracy reaches 40.85\%. 



Overall, input precision of 4-bit with ADC resolution of 8-bit provides an acceptable accuracy of 55.88\%, which is only 0.08\% lower than the maximum accuracy of 55.93\% when both parameters are set to 8, and is only 0.5\% lower than the accuracy of 56.15\% achieved on GPU with float32 first-layer precision.



\subsubsection{Put It All Together}
The maximum accuracy achieved by the three models exhibits a monotonic increase: 35.90\% for AlexNet, 53.84\% for ResNet, and 55.93\% for DenseNet. Although the values differ, the peak accuracy for each model is attained when the input precision and ADC resolution are set to their maximum values (8-bit). Overall, input precision of 4-bit with ADC resolution of 8-bit provides negligible accuracy loss compared to the same BNN model evaluated on GPU with float32 first-layer activation.


\begin{figure}[htbp]
  \centering
  \includegraphics[width=\linewidth]{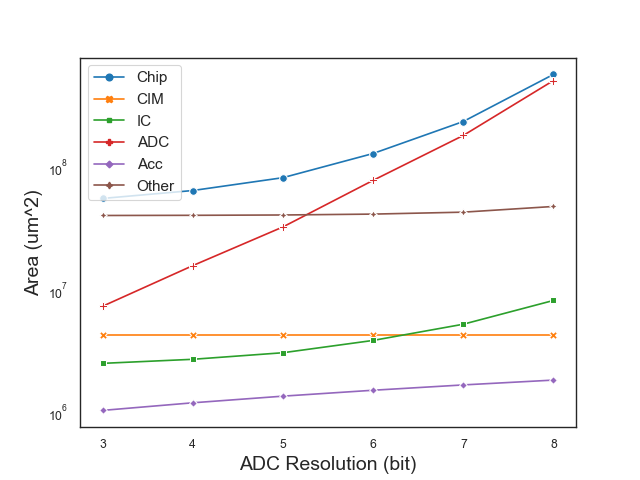}
  \caption{\textbf{Chip area breakdown for Binary AlexNet with different ADC resolutions} --- \texttt{\textit{Chip}} means the total chip area. \texttt{\textit{CIM, IC, ADC, Acc}} means CIM array area, interconnect circuits area, ADC area, and accumulation circuits area on chip respectively. \texttt{\textit{Other}} means other peripheries (e.g. decoders, mux, switch matrix, buffers, pooling, and activation units) on the chip. Note that the figure is in log scale.}
  \label{fig:hw_area_alexnet}
\end{figure}

\begin{table*}[htbp]
    \caption {\textbf{Chip Area breakdown for different BNN models with different ADC resolutions} --- Accumulation circuits include adders, shift/Adds on the subarray level, and includes accumulation units on the PE/Tile/Global level. Other peripheries include decoders, mux, switch matrix, buffers, pooling, activation units, etc.} 
   
    \label{tab:hw_area_together} 
    \begin{center}
    \resizebox{\textwidth}{!}{
        \input{Assets/hw/area/together.tex}
    }
    \end{center}
    \vspace{0.05in}
\end{table*}

\subsection{Area}

\subsubsection{AlexNet}

Figure \ref{fig:hw_area_alexnet} displays the chip area and its components as a function of ADC resolution for Binary AlexNet. It is evident that the total area of the ADC increases at the fastest rate as the ADC resolution increases, and this increase is the primary reason for the growth in the chip area. This is reasonable because higher ADC resolution necessitates a larger ADC circuit area. The total accumulation circuits on the chip, although having the lowest value among these components, increased by 76\% from 1.05 $mm^2$ to 1.85 $mm^2$. The other peripheries, which occupy most of the chip area when the ADC resolution is low, have only a small increase, resulting in a much smaller portion compared to the ADC area when the ADC resolution is 8. The CIM (Computing-in-Memory) array area, which is for the crossbar, remains exactly constant since it is not impacted by ADC circuits. As a result, the interconnect circuits (IC) area, although initially smaller than the CIM array, becomes larger than it when the ADC resolution is greater than 7.

\subsubsection{ResNet}

The total ADC area and chip area exhibit a similar trend as that of AlexNet, but with some notable differences. The total CIM array area remains constant, which is roughly half the size of AlexNet's CIM array (2.34 $mm^2$ vs. 4.32 $mm^2$), despite ResNet having more layers. This is due to AlexNet having three large dense layers (9216 * 4096, 4096 * 4096, 4096 * 1000) at the end, which requires more parameters to be stored in the CIM array. In contrast, ResNet only has one smaller dense layer (512 * 1000) at the end, resulting in fewer parameters to be stored in the CIM array. Consequently, the total IC area of ResNet is larger than the CIM array area from the start when ADC resolution is 3.


\subsubsection{DenseNet}

Despite having a more intricate and deeper network architecture than ResNet, DenseMet exhibits a larger CIM array area of 4.54 $mm^2$, which is almost twice the size of ResNet's 2.34 $mm^2$ and slightly larger than AlexNet's 4.32 $mm^2$. The total accumulation circuits and other periphery areas scale similar to the other models, yet the total IC area remains only marginally larger than that of ResNet across all ADC resolution levels (2.66 $mm^2$ vs. 2.52 $mm^2$ for ADC resolution = 3 and 8.66 $mm^2$ vs. 8.20 $mm^2$ for ADC resolution = 8). Consequently, the IC area is initially smaller than the CIM array area and surpasses it only when the ADC resolution exceeds 6.


\subsubsection{Put It All Together}

Table \ref{tab:hw_area_together} lists the precise values of chip area as well as its components as a function of ADC resolution for AlexNet, ResNet, and Densenet. Due to three large dense layers at the end of AlexNet, the chip area of AlexNet, although still smaller than DenseNet, is larger than ResNet, despite the fact that AlexNet is much shallower than the other two. The CIM array area shows a similar pattern, as ResNet is the smallest, and DenseNet is the largest. 

Total ADC area is the main factor of increase in chip area for all three models when ADC resolution grows. Similar for all three models, the ADC area grows exponentially (69x) with ADC resolution, and its portion on the chip grows from 13\% to 89\%. Total IC area, on the other hand, although increasing with ADC resolution, has a very similar value among the three models, which means the area of IC is almost independent of the model structure.

\begin{figure}[htbp]
  \centering
  \includegraphics[width=\linewidth]{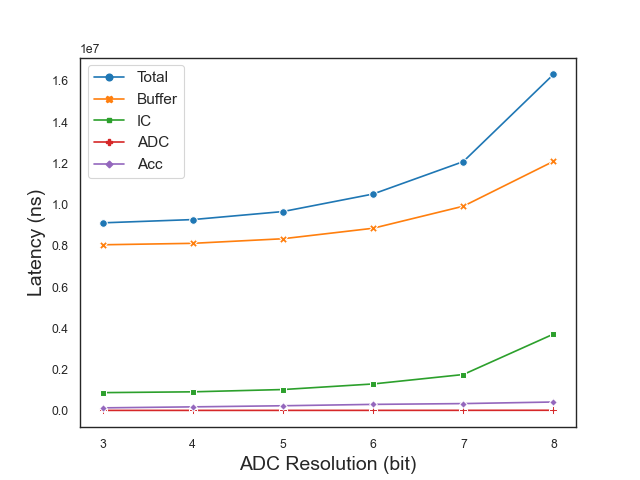}
  \caption{\textbf{Inference latency breakdown for Binary AlexNet with different ADC resolutions} --- \texttt{\textit{Total, Buffer, IC, ADC, Acc}} means total, buffer, interconnect circuits, ADC, and accumulation circuits read latency per image respectively.}
  \label{fig:hw_latency_alexnet}
\end{figure}

\begin{table*}[htbp]
    \caption {\textbf{Inference latency breakdown for different BNN models with different ADC resolutions.}} 
    \label{tab:hw_latency_together} 
    \begin{center}
    \resizebox{\textwidth}{!}{
        \input{Assets/hw/latency/together.tex}
    }
    \end{center}
    \vspace{0.05in}
\end{table*}

\begin{figure}[htbp]
  \centering
  \includegraphics[width=\linewidth]{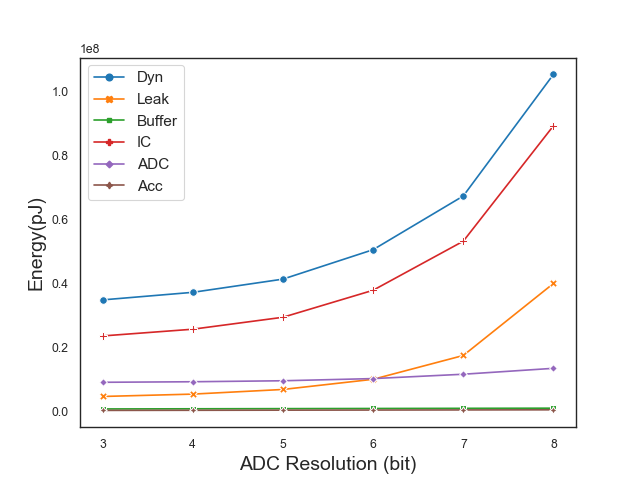}
  \caption{\textbf{Per-image energy breakdown for Binary AlexNet with different ADC resolutions} --- \texttt{\textit{Dyn}} means the total read dynamic energy. \texttt{\textit{Leak}} means the leakage energy. \texttt{\textit{Buffer, IC, ADC, Acc}} means buffer, interconnect circuits, ADC, and accumulation circuits read dynamic energy respectively.}
  \label{fig:hw_power_alexnet}
\end{figure}

\begin{table*}[htbp]
    \caption {\textbf{Per-image energy breakdown for different BNN models with different ADC resolutions.}} 
    \label{tab:hw_power_together} 
    \begin{center}
    \resizebox{\textwidth}{!}{
        \input{Assets/hw/power/together.tex}
    }
    \end{center}
    \vspace{0.05in}
\end{table*}

\subsection{Latency}

\subsubsection{AlexNet}

The per-image inference latency and their components as functions of ADC resolution for Binary AlexNet are shown in Figure \ref{fig:hw_latency_alexnet}. The main contributor to the total inference latency is the buffer read latency, which increases from 8.05 \textit{ms} to 12.1 \textit{ms}, resulting in a similar increasing trend in the overall latency.


 Although the buffer read latency is larger than 8 ms even at the beginning, the second major component, IC read latency, is smaller by one order of magnitude. The ADC read latency is the least significant component of the total latency, with a magnitude level of 10 us. Although it also increases with ADC resolution, its increase ratio (156.25\%) is smaller than the previously mentioned components.

\subsubsection{ResNet}

Surprisingly, despite ResNet's deeper model, the total per-image latency is even smaller than that of AlexNet. This is due to ResNet having a much smaller dense layer, resulting in a lower inference latency. Similar to AlexNet, the buffer read latency is the major contributor to the whole latency. The ADC read latency, although keeping increasing with ADC resolution, is one order of magnitude less than buffer read latency and is also the least contributor to the total inference latency.



\subsubsection{DenseNet}

The total inference latency, which ranges from 2.11 \textit{ms} when ADC resolution is 3 to 3.56 \textit{ms} when ADC resolution is 8, is of similar magnitude as ResNet and significantly smaller than AlexNet. Unlike the case for chip area, where the area of DenseNet is comparable to AlexNet and larger than ResNet, this implies that latency is less related to the depth of the model but more to the size of weights. The existence of large dense layers in AlexNet may be the primary reason for its high latency. Due to its dense-connected nature, the IC read latency of DenseNet is over twice the value of ResNet. On the other hand, the ADC read latency is the same as ResNet, which could be attributed to the similarity in the internal mapping of layers.


\subsubsection{Put It All Together}

Table \ref{tab:hw_latency_together} presents the precise values of the clock cycle, per-image inference latency, and its components as functions of ADC resolution for AlexNet, ResNet, and DenseNet. The clock cycle is exactly the same for all three models, regardless of the ADC resolution used, indicating its independence of model structures.

AlexNet has the highest total latency among the three models and also has the highest latency for all its components. This finding suggests that the presence of large dense layers is one of the main contributors to latency in AlexNet.

Across all ADC resolutions, the main component of total inference latency for all three models is the buffer read latency with the ADC read latency remaining the least significant component in all three models. Due to the pipelined architecture, deep and complex model structure has little impact on latency: DenseNet's latency is only 1.04x of ResNet.

\begin{figure}[htbp]
  \centering
  \includegraphics[width=\linewidth]{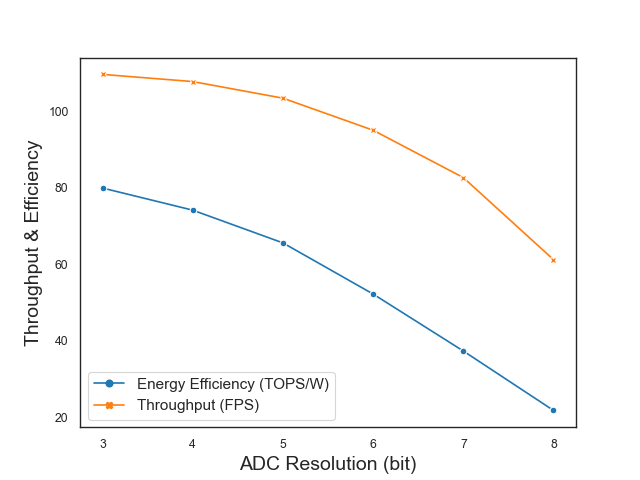}
  \caption{\textbf{Throughput and energy efficiency for Binary Alexnet with different ADC resolutions.}}
  \label{fig:hw_efficiency_alexnet}
\end{figure}

\begin{table*}[htbp]
    \centering
    \caption {\textbf{Throughput and efficiency of the simulated NVM accelerators for different BNN models with different ADC resolutions.}} 
    \label{tab:hw_efficiency_together} 
    \resizebox{0.8\textwidth}{!}{
        \input{Assets/hw/efficiency/together.tex}
    }
\end{table*}

\subsection{Energy}

\subsubsection{AlexNet}

Figure \ref{fig:hw_power_alexnet} displays the dynamic energy consumption of Binary AlexNet, including dynamic energies and leakage energy as functions of ADC resolution. As the ADC resolution increases, the total read dynamic energy and leakage energy both grow rapidly. 
The IC read dynamic energy is the main contributor to the total dynamic energy consumption. Despite the fast increase in ADC resolution, the ADC read dynamic energy grows at a relatively slower rate (from 9.17 \textit{uJ} to 13.54 \textit{uJ}, an increase of approximately 47\%), and its overall value is relatively small. This results in a cross-point of leakage energy and ADC read dynamic energy: the latter is higher until ADC resolution reaches 7, after which the former is higher. The buffer read energy and accumulation circuit energy are the two least significant components of the total energy, at the level of \textit{uJ}, while the total energy can be tens to hundreds of times larger.

\subsubsection{ResNet}

Although the total read dynamic energy is at a similar level to that of AlexNet, the leakage energy of ResNet is significantly smaller. When ADC resolution is 3, the leakage energy for ResNet is only 10.7\% of that of AlexNet, and this ratio decreases to 10.2\% when ADC resolution is 8. Looking further at the latency and area of ResNet, an interesting conclusion can be drawn: while a larger chip area leads to larger leakage power, the longer inference latency is a more significant factor in the higher total leakage energy consumption of AlexNet. In contrast, the ADC read dynamic latency of ResNet is higher than that of AlexNet for all ADC resolutions, with a faster increase speed: 190\%. The IC read dynamic energy is still the main factor contributing to the total energy consumption.


\subsubsection{DenseNet}

Although the total read dynamic energy of DenseNet is the largest among the three models, its leakage energy is greater than ResNet but less than AlexNet. On the other hand, the IC read dynamic energy, which is almost twice as large as its AlexNet and ResNet counterparts, is the major contributor to the large total dynamic energy. The ADC read dynamic energy, although about two times larger than ResNet, has a similar increasing speed. The accumulation circuits read energy and buffer read energy are the two least significant components of the total dynamic energy.


\subsubsection{Put It All Together}

Table \ref{tab:hw_power_together} provides precise values of chip read dynamic energy, leakage energy, and power, as well as other read energies for AlexNet, ResNet, and DenseNet at different ADC resolutions. DenseNet has the largest total read dynamic energy, which is twice the value of the other two models. This similar ratio also holds true for the IC read dynamic energy, the main contributor to the total dynamic energy.

In contrast, AlexNet has the highest leakage energy for all ADC resolutions, which is contributed by the long inference latency of AlexNet. Across all three models, the buffer read energy and accumulation circuits read energy have the least significant contribution to the total dynamic energy.

\subsection{Throughput \& Efficiency}

Figure \ref{fig:hw_efficiency_alexnet} shows the energy efficiency and throughput as functions of ADC resolution for Binary Alexnet. Table \ref{tab:hw_efficiency_together} provides precise values of these metrics. The figure distinctly illustrates a significant decrease in throughput with increasing ADC resolution: the throughput at an ADC resolution of 8-bits is merely 55\% of that at a 1-bit resolution. Additionally, the decline in energy efficiency is even more pronounced. An interesting observation is that AlexNet has the lowest TOPS and FPS throughput, which is likely due to its large per-image latency.

ResNet has the highest FPS throughput (a little higher than DenseNet), possibly due to its simpler network structure compared to DenseNet. ResNet also has the highest TOPS/W energy efficiency, which is more than 50\% higher than the other two models. With regards to the TOPS/$mm^2$ compute efficiency, despite its large TOPS throughput, AlexNet is significantly lower (about 10x) than the other two models. Overall, as ADC resolution increases, all four metrics decrease since more chip area, latency, and power are needed to perform more accurate ADC.



%% file: Assets/acc/alexnet.tex

    
    
    

\renewcommand\arraystretch{1.05}
\Large
\begin{tabular}{c|cccccc}
    \toprule
    \multirow{2}{4em}{\centering \textbf{Input Precision}} &\multicolumn{6}{c}{\textbf{ADC Resolution}} \\
    ~ & {3} & {4} & {5}& {6}& {7} & {8}  \\ 
    \midrule
    {1} & 0.17 & 1.43 & 3.64 & 6.07 & 7.67 & 7.55 \\ 
    {2} & 0.17 & 3.03 & 9.27 & 22.58 & 29.97 & 30.71 \\ 
    {3} & 0.21 & 4.35 & 13.81 & 29.30 & 34.84 & 35.60 \\ 
    {4} & 0.16 & 4.95 & 15.51 & 30.76 & 35.57 & 35.76 \\ 
    {5} & 0.15 & 5.18 & 15.96 & 31.06 & 35.52 & 36.11 \\ 
    {6} & 0.18 & 5.24 & 16.00 & 30.99 & 35.54 & 35.84 \\ 
    {7} & 0.29 & 5.31 & 16.18 & 31.28 & 35.43 & 35.90 \\ 
    {8} & 0.31 & 5.31 & 16.31 & 31.22 & 35.56 & 35.90 \\
    \bottomrule
\end{tabular}

%% file: Assets/acc/resnet.tex




\renewcommand\arraystretch{1.05}
\Large
\begin{tabular}{c|cccccc}
    \toprule
    \multirow{2}{4em}{\centering \textbf{Input Precision}} &\multicolumn{6}{c}{\textbf{ADC Resolution}} \\
    ~ & {3} & {4} & {5}& {6}& {7} & {8}  \\ 
    \midrule
    {1} & 0.35 & 2.05 & 4.23 & 6.96 & 6.46 & 6.71 \\ 
    {2} & 0.60 & 15.23 & 18.82 & 38.60 & 36.08 & 39.72 \\ 
    {3} & 0.55 & 22.35 & 24.17 & 45.86 & 48.36 & 51.81 \\ 
    {4} & 0.52 & 25.08 & 25.67 & 46.79 & 51.08 & 53.61 \\ 
    {5} & 0.70 & 26.04 & 26.04 & 46.89 & 51.60 & 53.98 \\ 
    {6} & 0.79 & 26.32 & 26.61 & 47.19 & 51.93 & 53.81 \\ 
    {7} & 0.65 & 26.51 & 26.68 & 47.08 & 52.02 & 53.82 \\ 
    {8} & 0.77 & 26.58 & 26.42 & 47.00 & 51.88 & 53.84 \\
    \bottomrule
\end{tabular}

%% file: Assets/acc/densenet.tex




\renewcommand\arraystretch{1.05}
\Large
\begin{tabular}{c|cccccc}
    \toprule
    \multirow{2}{4em}{\centering \textbf{Input Precision}} &\multicolumn{6}{c}{\textbf{ADC Resolution}} \\
    ~ & {3} & {4} & {5}& {6}& {7} & {8}  \\ 
    \midrule
    {1} & 0.24 & 1.89 & 3.34 & 5.76 & 5.14 & 5.40 \\ 
    {2} & 0.75 & 22.65 & 20.64 & 32.20 & 40.67 & 40.85 \\ 
    {3} & 0.91 & 32.98 & 26.39 & 42.91 & 54.05 & 54.15 \\ 
    {4} & 0.84 & 34.82 & 27.04 & 45.49 & 55.42 & 55.88 \\ 
    {5} & 0.85 & 35.22 & 27.56 & 46.19 & 55.52 & 56.22 \\ 
    {6} & 1.04 & 35.41 & 27.78 & 46.58 & 55.44 & 56.18 \\ 
    {7} & 0.85 & 36.07 & 27.47 & 46.49 & 55.58 & 56.04 \\ 
    {8} & 0.95 & 35.92 & 27.72 & 46.37 & 55.68 & 55.93 \\
    \bottomrule
\end{tabular}

%% file: Assets/hw/area/together.tex
\newcommand{\tabincell}[2]{\begin{tabular}{@{}#1@{}}#2\end{tabular}}
\renewcommand\arraystretch{1.3}
\Large
\begin{tabular}
{cccccccc}
    \toprule
    \textbf{\tabincell{c}{BNN \\ model}} & \textbf{\tabincell{c}{ADC \\ resolution}} & \textbf{\tabincell{c}{Chip area \\ ($mm^2$)}} & \textbf{\tabincell{c}{Total CIM array \\  ($mm^2$)}} & \textbf{\tabincell{c}{Total IC area \\ ($mm^2$)}} & \textbf{\tabincell{c}{Total ADC area \\ ($mm^2$)}} & \textbf{\tabincell{c}{Total accumulation \\ circuits ($mm^2$)}} & \textbf{\tabincell{c}{Other peripheries \\ ($mm^2$)}} \\ 

    \midrule
    {AlexNet} & {3} & 55.97 & 4.32 & 2.54 & 7.40 & 1.05 & 40.65 \\ 
    {} & {4} & 64.92 & 4.32 & 2.74 & 15.87 & 1.21 & 40.78 \\ 
    {} & {5} & 82.62 & 4.32 & 3.10 & 32.79 & 1.37 & 41.04 \\ 
    {} & {6} & 130.04 & 4.32 & 3.90 & 78.56 & 1.53 & 41.72 \\ 
    {} & {7} & 236.97 & 4.32 & 5.29 & 182.39 & 1.69 & 43.27 \\ 
    {} & {8} & 573.53 & 4.32 & 8.26 & 510.94 & 1.85 & 48.15 \\ 
    \midrule
    {ResNet} & {3} & 32.73 & 2.34 & 2.52 & 4.02 & 0.57 & 23.28 \\ 
    {} & {4} & 37.67 & 2.34 & 2.72 & 8.60 & 0.65 & 23.35 \\ 
    {}  & {5} & 47.43 & 2.34 & 3.08 & 17.78 & 0.74 & 23.49 \\ 
    {} & {6} & 73.51 & 2.34 & 3.87 & 42.60 & 0.83 & 23.86 \\ 
    {} & {7} & 132.11 & 2.34 & 5.25 & 98.90 & 0.91 & 24.70 \\ 
    {} & {8} & 315.95 & 2.34 & 8.20 & 277.06 & 1.00 & 27.35 \\
    \midrule
    {DenseNet} & {3} & 62.73 & 4.54 & 2.66 & 7.77 & 1.08 & 46.68 \\ 
    {} & {4} & 72.13 & 4.54 & 2.87 & 16.65 & 1.25 & 46.82 \\ 
    {}  & {5} & 90.70 & 4.54 & 3.25 & 34.41 & 1.41 & 47.09 \\ 
    {} & {6} & 140.45 & 4.54 & 4.09 & 82.43 & 1.58 & 47.81 \\ 
    {} & {7} & 252.65 & 4.54 & 5.55 & 191.38 & 1.75 & 49.43 \\ 
    {} &{8} & 605.80 & 4.54 & 8.66 & 536.12 & 1.91 & 54.56 \\
    \bottomrule
\end{tabular}

%% file: Assets/hw/latency/together.tex
\newcommand{\tabincell}[2]{\begin{tabular}{@{}#1@{}}#2\end{tabular}}
\renewcommand\arraystretch{1.4}
\Large
\begin{tabular}{ccccccccc}   
    \toprule
    \textbf{\tabincell{c}{BNN \\ model}} &
    \textbf{\tabincell{c}{ADC \\ resolution}} &
    \textbf{\tabincell{c}{Chip clock \\ period (ns)}} &
    \textbf{\tabincell{c}{Chip clock cycle \\ (per image) (ms)}} & \textbf{\tabincell{c}{Chip buffer latency \\(per image) (ms)}} & \textbf{\tabincell{c}{Chip IC latency \\ (per image) (ms)}} & 
    \textbf{\tabincell{c}{ADC latency \\ (us)}} & \textbf{\tabincell{c}{Accumulation circuits \\ latency (us)}} & 
    \\
    \midrule
    {{AlexNet}} & {3} & 1.96 & 9.11 & 8.05 & 0.87 & 11.22 & 134.66 \\ 
    {} &{4} & 2.00 & 9.27 & 8.12 & 0.91 & 11.43 & 182.88 \\ 
    {} &{5} & 2.07 & 9.66 & 8.34 & 1.02 & 11.84 & 236.73 \\ 
    {} &{6} & 2.21 & 10.51 & 8.85 & 1.29 & 12.64 & 303.39 \\ 
    {} &{7} & 2.49 & 12.09 & 9.92 & 1.75 & 14.24 & 341.90 \\ 
    {} &{8} & 3.05 & 16.32 & 12.10 & 3.71 & 17.45 & 418.86 \\
    \midrule
    {{ResNet}} &{3} & 1.96 & 2.02 & 1.86 & 0.12 & 5.72 & 60.02 \\ 
    {} &{4} & 2.00 & 2.05 & 1.87 & 0.13 & 5.82 & 61.14 \\ 
    {} &{5} & 2.07 & 2.13 & 1.92 & 0.15 & 6.03 & 66.32 \\ 
    {} &{6} & 2.21 & 2.29 & 2.04 & 0.18 & 6.44 & 70.83 \\ 
    {} &{7} & 2.49 & 2.60 & 2.28 & 0.24 & 7.26 & 79.82 \\ 
    {} &{8} & 3.05 & 3.41 & 2.78 & 0.53 & 8.89 & 97.79 \\ 
    \midrule
    {{DenseNet}} &{3} & 1.96 & 2.11 & 1.95 & 0.27 & 5.72 & 68.60 \\ 
    {} &{4} & 2.00 & 2.15 & 1.97 & 0.28 & 5.82 & 72.78 \\ 
    {} &{5} & 2.07 & 2.23 & 2.02 & 0.32 & 6.03 & 78.38 \\ 
    {} &{6} & 2.21 & 2.39 & 2.14 & 0.44 & 6.44 & 83.71 \\ 
    {} &{7} & 2.49 & 2.72 & 2.40 & 0.65 & 7.26 & 94.34 \\ 
    {} &{8} & 3.05 & 3.56 & 2.93 & 1.24 & 8.89 & 115.57 \\ \bottomrule
\end{tabular}

%% file: Assets/hw/power/together.tex
\newcommand{\tabincell}[2]{\begin{tabular}{@{}#1@{}}#2\end{tabular}}
\renewcommand\arraystretch{1.6}
\huge
\begin{tabular}{ccccccccc}   
    \toprule
     \textbf{\tabincell{c}{BNN \\ model}} 
     & \textbf{\tabincell{c}{ADC \\ resolution}} 
     & \textbf{\tabincell{c}{Dynamic energy \\(per image) (uJ)}}
     & \textbf{\tabincell{c}{Leakage energy \\(per image) (uJ)}} 
     & \textbf{\tabincell{c}{Buffer dynamic energy \\ (per image) (uJ)}} 
     & \textbf{\tabincell{c}{IC dynamic energy \\(per image) (uJ)}}
     & \textbf{\tabincell{c}{ADC dynamic \\ energy (uJ)}} 
     & \textbf{\tabincell{c}{Accumulation circuits \\ dynamic energy (uJ)}}
     \\
     \midrule
    {AlexNet} & {3} & 34.89 & 4.75 & 0.86 & 23.64 & 9.17 & 0.40 \\ 
    {} &{4} & 37.25 & 5.48 & 0.90 & 25.74 & 9.34 & 0.44 \\ 
    {} &{5} & 41.37 & 6.93  & 0.95 & 29.47 & 9.65 & 0.47 \\ 
    {} &{6} & 50.56 & 10.11 & 1.00 & 37.90 & 10.33 & 0.51 \\ 
    {} &{7} & 67.32 & 17.54 & 1.05 & 53.21 & 11.69 & 0.55 \\ 
    {} &{8} & 105.27 & 40.02 & 1.10 & 89.23 & 13.54 & 0.58 \\
    \midrule
    {ResNet} &{3} & 32.27 & 0.51  & 0.83 & 18.39 & 10.96 & 0.53 \\ 
    {} &{4} & 34.53 & 0.59  & 0.89 & 20.19 & 11.29 & 0.61 \\ 
    {} &{5} & 38.48 & 0.74  & 0.95 & 23.37 & 11.92 & 0.69 \\ 
    {} &{6} & 47.17 & 1.06  & 1.02 & 30.61 & 13.21 & 0.76 \\ 
    {} &{7} & 63.46 & 1.83  & 1.08 & 44.21 & 15.78 & 0.84 \\ 
    {} &{8} & 102.32 & 4.06  & 1.14 & 77.79 & 20.91 & 0.92 \\
    \midrule
    {DenseNet} &{3} & 107.23 & 0.93  & 2.20 & 65.28 & 32.02 & 2.33 \\ 
    {} &{4} & 114.15 & 1.07  & 2.32 & 70.97 & 32.80 & 2.68 \\ 
    {} &{5} & 126.29 & 1.35  & 2.43 & 81.11 & 34.34 & 3.03 \\ 
    {} &{6} & 152.83 & 1.95  & 2.54 & 104.10 & 37.42 & 3.38 \\ 
    {} &{7} & 201.22 & 3.35 & 2.66 & 145.87 & 43.58 & 3.73 \\ 
    {} &{8} & 312.23 & 7.40  & 2.77 & 244.09 & 55.91 & 4.08 \\
    \bottomrule
\end{tabular}

%% file: Assets/hw/efficiency/together.tex
\newcommand{\tabincell}[2]{\begin{tabular}{@{}#1@{}}#2\end{tabular}}
\renewcommand\arraystretch{1.1}
\normalsize
\begin{tabular}{cccccc} 
    \toprule
    \textbf{\tabincell{c}{BNN \\ model}} & \textbf{\tabincell{c}{ADC \\resolution}} & \tabincell{c}{\textbf{Energy efficiency} \\ (TOPS/W)} & \tabincell{c}{\textbf{Throughput} \\ (TOPS)} & \tabincell{c}{\textbf{Throughput} \\ (FPS)} & \tabincell{c}{\textbf{Compute efficiency} \\ (TOPS/$cm^2$)} \\
    \midrule
    {AlexNet} & {3} & 79.99 & 0.424 & 109.71 & 0.76 \\ 
    {} &{4} & 74.20 & 0.417 & 107.84 & 0.64 \\ 
    {} &{5} & 65.64 & 0.400 & 103.49 & 0.48 \\ 
    {} &{6} & 52.26 & 0.368 & 95.11 & 0.28 \\ 
    {} &{7} & 37.37 & 0.320 & 82.73 & 0.14 \\ 
    {} &{8} & 21.82 & 0.237 & 61.28 & 0.04 \\
    \midrule
    {ResNet} &{3} & 137.37 & 2.72 & 495.46 & 8.31 \\ 
    {} &{4} & 128.20 & 2.68 & 487.11 & 7.10 \\
    {} &{5} & 114.78 & 2.58 & 468.89 & 5.43 \\ 
    {} &{6} & 93.34 & 2.40 & 437.28 & 3.27 \\ 
    {} &{7} & 68.96 & 2.11 & 385.07 & 1.60 \\ 
    {} &{8} & 42.32 & 1.61 & 293.32 & 0.51 \\
    \midrule
    {DenseNet} &{3} & 62.81 & 3.92 & 473.48 & 6.26 \\ 
    {} &{4} & 58.96 & 3.86 & 465.47 & 5.35 \\ 
    {} &{5} & 53.22 & 3.71 & 448.09 & 4.09 \\ 
    {} &{6} & 43.89 & 3.46 & 417.85 & 2.47 \\ 
    {} &{7} & 33.21 & 3.05 & 367.96 & 1.21 \\ 
    {} &{8} & 21.25 & 2.33 & 280.96 & 0.38 \\
    \bottomrule
    \end{tabular}

%% file: Sections/conclusion.tex
\section{Conclusions and Future Work}
\label{sec:conclusion}

In this paper, we benchmark the inference accuracy, chip area, latency, power, throughput, and efficiency of three modern BNN models: BAlexnet, BResnet18, and BDensenet28, which are trained and evaluated on Imagenet and deployed on NeuroSim, a NVM crossbar-based PIM simulator. Our work shows that increasing ADC resolution and input precision generally leads to higher accuracy but lower throughput, and using 4-bit input precision and 8-bit ADC resolution provides negligible accuracy loss compared to using float32 first-layer activation on GPU. Additionally, although the ADC area grows exponentially as its resolution increases and dominates the chip area, the buffer latency and the interconnection dynamic energy are still the bottlenecks of latency and energy, respectively. Finally, as the pipeline architecture of the NVM accelerator significantly reduces latency for deeper models, large fully-connected layers become the bottleneck of chip performance, especially the latency. Being the first comprehensive benchmark of modern BNN on ImageNet with NVM crossbar-based PIM devices, we believe this work could provide very detailed raw and analyzed data as well as a clearer guideline for the future co-design of BNN models and PIM accelerators.

    


Our future work entails exploring state-of-the-art BNN models, such as the ones proposed by Guo et al. \cite{guo2022join}, Zhang et al. \cite{zhang2023binarized}, Liu et al. \cite{liu2020reactnet}, and Zhang et al. \cite{Zhang_2022_CVPR}, to address the limited accuracy observed in our current models when applied to complex tasks and datasets like ImageNet. We aim to assess their performance and appropriateness for NVM crossbar architectures, taking into account factors such as accuracy, robustness, and chip performance. This exploration will further our understanding of BNN models and contribute to the development of more efficient and effective PIM accelerators.

%% file: Sections/acknowledgement.tex
\section*{Acknowledgement}
This work was supported in part by the NSF Award \#2007832 and a research gift from Qualcomm.

%% file: main.bbl
\begin{thebibliography}{10}
\providecommand{\url}[1]{#1}
\csname url@samestyle\endcsname
\providecommand{\newblock}{\relax}
\providecommand{\bibinfo}[2]{#2}
\providecommand{\BIBentrySTDinterwordspacing}{\spaceskip=0pt\relax}
\providecommand{\BIBentryALTinterwordstretchfactor}{4}
\providecommand{\BIBentryALTinterwordspacing}{\spaceskip=\fontdimen2\font plus
\BIBentryALTinterwordstretchfactor\fontdimen3\font minus
  \fontdimen4\font\relax}
\providecommand{\BIBforeignlanguage}[2]{{%
\expandafter\ifx\csname l@#1\endcsname\relax
\typeout{** WARNING: IEEEtran.bst: No hyphenation pattern has been}%
\typeout{** loaded for the language `#1'. Using the pattern for}%
\typeout{** the default language instead.}%
\else
\language=\csname l@#1\endcsname
\fi
#2}}
\providecommand{\BIBdecl}{\relax}
\BIBdecl

\bibitem{shafiee_isaac_2016}
A.~Shafiee, A.~Nag, N.~Muralimanohar, R.~Balasubramonian, J.~P. Strachan,
  M.~Hu, R.~S. Williams, and V.~Srikumar, ``{{ISAAC}: A Convolutional Neural
  Network Accelerator with In-Situ Analog Arithmetic in Crossbars},''
  \emph{Int'l Symp. on Computer Architecture (ISCA)}, 2016.

\bibitem{ankit_puma_2019}
A.~Ankit, I.~E. Hajj, S.~R. Chalamalasetti, G.~Ndu, M.~Foltin, R.~S. Williams,
  P.~Faraboschi, W.-m.~W. Hwu, J.~P. Strachan, K.~Roy, and D.~S. Milojicic,
  ``{{PUMA}: A Programmable Ultra-efficient Memristor-based Accelerator for
  Machine Learning Inference},'' \emph{Int'l Conf. on Architectural Support for
  Programming Languages and Operating Systems (ASPLOS)}, 2019.

\bibitem{ankit2020panther}
A.~Ankit, I.~El~Hajj, S.~R. Chalamalasetti, S.~Agarwal, M.~Marinella,
  M.~Foltin, J.~P. Strachan, D.~Milojicic, W.-M. Hwu, and K.~Roy, ``{{Panther}:
  A Programmable Architecture for Neural Network Training Harnessing
  Energy-Efficient ReRAM},'' \emph{IEEE Trans. on Computers (TC)}, vol.~69,
  no.~8, pp. 1128--1142, 2020.

\bibitem{boybat_neuromorphic_2018}
I.~Boybat, M.~Le~Gallo, S.~R. Nandakumar, T.~Moraitis, T.~Parnell, T.~Tuma,
  B.~Rajendran, Y.~Leblebici, A.~Sebastian, and E.~Eleftheriou, ``{Neuromorphic
  computing with multi-memristive synapses},'' \emph{Nature Communications},
  vol.~9, no.~1, p. 2514, 2018.

\bibitem{hung2021challenges}
J.-M. Hung, C.-J. Jhang, P.-C. Wu, Y.-C. Chiu, and M.-F. Chang, ``{Challenges
  and Trends of Nonvolatile In-Memory-Computation Circuits for AI Edge
  Devices},'' \emph{IEEE Open Journal of the Solid-State Circuits Society},
  vol.~1, pp. 171--183, 2021.

\bibitem{yu2018neuro}
S.~Yu, ``{Neuro-Inspired Computing with Emerging Nonvolatile Memorys},''
  \emph{Proceedings of the IEEE}, vol. 106, no.~2, pp. 260--285, 2018.

\bibitem{chakraborty2020resistive}
I.~Chakraborty, M.~Ali, A.~Ankit, S.~Jain, S.~Roy, S.~Sridharan, A.~Agrawal,
  A.~Raghunathan, and K.~Roy, ``{Resistive Crossbars as Approximate Hardware
  Building Blocks for Machine Learning: Opportunities and Challenges},''
  \emph{Proceedings of the IEEE}, vol. 108, no.~12, pp. 2276--2310, 2020.

\bibitem{zhang_fracbnn_2020}
Y.~Zhang, J.~Pan, X.~Liu, H.~Chen, D.~Chen, and Z.~Zhang, ``{FracBNN: Accurate
  and FPGA-Efficient Binary Neural Networks with Fractional Activations},''
  \emph{Int'l Symp. on Field-Programmable Gate Arrays (FPGA)}, 2021.

\bibitem{hubara_binarized_nodate}
I.~Hubara, M.~Courbariaux, D.~Soudry, R.~El-Yaniv, and Y.~Bengio, ``{Binarized
  Neural Networks},'' \emph{Conf. on Neural Information Processing Systems
  ({NeurIPS})}, 2016.

\bibitem{li_bcnn_2021}
Y.~Li, T.~Geng, A.~Li, and H.~Yu, ``{BCNN: Binary Complex Neural Network},''
  \emph{Microprocessors and Microsystems}, vol.~87, p. 104359, 2021.

\bibitem{qin_binary_2020}
H.~Qin, R.~Gong, X.~Liu, X.~Bai, J.~Song, and N.~Sebe, ``{Binary Neural
  Networks: A Survey},'' \emph{Pattern Recognition}, vol. 105, p. 107281, 2020.

\bibitem{liang_fp-bnn_2018}
S.~Liang, S.~Yin, L.~Liu, W.~Luk, and S.~Wei, ``{FP-BNN: Binarized Neural
  Network on FPGA},'' \emph{Neurocomputing}, vol. 275, pp. 1072--1086, 2018.

\bibitem{lin_towards_nodate}
X.~Lin, C.~Zhao, and W.~Pan, ``{Towards Accurate Binary Convolutional Neural
  Network},'' \emph{Conf. on Neural Information Processing Systems
  ({NeurIPS})}, 2017.

\bibitem{crosssim}
\BIBentryALTinterwordspacing
S.~Agarwal, S.~J. Plimpton, R.~K. Schiek, I.~Richter, A.~H. Hsia, D.~R.
  Hughart, R.~B. Jacobs-Gedrim, C.~D. James, and M.~J. Marinella. (2017)
  Crosssim. [Online]. Available: \url{https://cross-sim.sandia.gov/}
\BIBentrySTDinterwordspacing

\bibitem{He2019NIA}
Z.~He, J.~Lin, R.~Ewetz, J.-S. Yuan, and D.~Fan, ``{Noise Injection Adaption:
  End-to-End ReRAM Crossbar Non-ideal Effect Adaption for Neural Network
  Mapping},'' \emph{Design Automation Conf. (DAC)}, 2019.

\bibitem{jain2020rxnn}
S.~Jain, A.~Sengupta, K.~Roy, and A.~Raghunathan, ``{RxNN: A Framework for
  Evaluating Deep Neural Networks on Resistive Crossbars},'' \emph{IEEE Trans.
  on Computer-Aided Design of Integrated Circuits and Systems (TCAD)}, vol.~40,
  no.~2, pp. 326--338, 2020.

\bibitem{roy2021txsim}
S.~Roy, S.~Sridharan, S.~Jain, and A.~Raghunathan, ``{Txsim: Modeling Training
  of Deep Neural Networks on Resistive Crossbar Systems},'' \emph{IEEE Trans.
  on Very Large-Scale Integration Systems (TVLSI)}, vol.~29, no.~4, pp.
  730--738, 2021.

\bibitem{zhu2020mnsim}
Z.~Zhu, H.~Sun, K.~Qiu, L.~Xia, G.~Krishnan, G.~Dai, D.~Niu, X.~Chen, X.~S. Hu,
  Y.~Cao \emph{et~al.}, ``{MNSIM 2.0: A Behavior-Level Modeling Tool for
  Memristor-based Neuromorphic Computing Systems},'' \emph{Great Lakes
  Symposium on VLSI}, pp. 83--88, 2020.

\bibitem{peng2019dnn+}
X.~Peng, S.~Huang, Y.~Luo, X.~Sun, and S.~Yu, ``{DNN+ NeuroSim: An End-to-End
  Benchmarking Framework for Compute-in-Memory Accelerators with Versatile
  Device Technologies},'' \emph{IEEE Int'l electron devices meeting (IEDM)},
  2019.

\bibitem{sun_computing--memory_2018}
X.~Sun, R.~Liu, X.~Peng, and S.~Yu, ``{Computing-in-Memory with SRAM and RRAM
  for Binary Neural Networks},'' \emph{IEEE Int'l Conf. on Solid-State and
  Integrated Circuit Technology (ICSICT)}, 2018.

\bibitem{zahedi_bcim_2022}
M.~Zahedi, T.~Shahroodi, S.~Wong, and S.~Hamdioui, ``{BCIM: Efficient
  Implementation of Binary Neural Network Based on Computation in Memory},''
  \emph{arXiv preprint arXiv:2211.06261}, 2022.

\bibitem{sun_fully_2018}
X.~Sun, X.~Peng, P.-Y. Chen, R.~Liu, J.-s. Seo, and S.~Yu, ``{Fully Parallel
  RRAM Synaptic Array for Implementing Binary Neural Network with (+ 1,- 1)
  Weights and (+ 1, 0) Neurons},'' \emph{Asia and South Pacific Design
  Automation Conf. (ASP-DAC)}, 2018.

\bibitem{sun_xnor-rram_2018}
X.~Sun, S.~Yin, X.~Peng, R.~Liu, J.-s. Seo, and S.~Yu, ``{XNOR-RRAM: A Scalable
  and Parallel Resistive Synaptic Architecture for Binary Neural Networks},''
  \emph{Design, Automation, and Test in Europe (DATE)}, 2018.

\bibitem{yin_xnor-sram_2020}
S.~Yin, Z.~Jiang, J.-S. Seo, and M.~Seok, ``{XNOR-SRAM: In-Memory Computing
  SRAM Macro for Binary/Ternary Deep Neural Networks},'' \emph{IEEE Journal of
  Solid-State Circuits}, vol.~55, no.~6, pp. 1733--1743, 2020.

\bibitem{kim_area-efficient_2019}
J.~Kim, J.~Koo, T.~Kim, Y.~Kim, H.~Kim, S.~Yoo, and J.-J. Kim,
  ``{Area-Efficient and Variation-Tolerant In-Memory BNN Computing using 6T
  SRAM Array},'' \emph{Symposium on VLSI Circuits}, pp. C118--C119, 2019.

\bibitem{schuman_survey_2017}
C.~D. Schuman, T.~E. Potok, R.~M. Patton, J.~D. Birdwell, M.~E. Dean, G.~S.
  Rose, and J.~S. Plank, ``{A survey of Neuromorphic Computing and Neural
  Networks in Hardware},'' \emph{arXiv preprint arXiv:1705.06963}, 2017.

\bibitem{aly_modern_2023}
O.~Mutlu, S.~Ghose, J.~Gómez-Luna, and R.~Ausavarungnirun, \emph{{A Modern
  Primer on Processing in Memory}}, M.~M.~S. Aly and A.~Chattopadhyay,
  Eds.\hskip 1em plus 0.5em minus 0.4em\relax Springer Nature Singapore, 2023,
  {Series Title: Computer Architecture and Design Methodologies}.

\bibitem{caminal_cape_2021}
H.~Caminal, K.~Yang, S.~Srinivasa, A.~K. Ramanathan, K.~Al-Hawaj, T.~Wu,
  V.~Narayanan, C.~Batten, and J.~F. Mart{\'\i}nez, ``{CAPE: A
  Content-Addressable Processing Engine},'' \emph{Int'l Symp. on
  High-Performance Computer Architecture (HPCA)}, 2021.

\bibitem{boukhobza2017emerging}
J.~Boukhobza, S.~Rubini, R.~Chen, and Z.~Shao, ``{Emerging NVM: A survey on
  Architectural Integration and Research Challenges},'' \emph{ACM Trans. on
  Design Automation of Electronic Systems (TODAES)}, vol.~23, no.~2, pp. 1--32,
  2017.

\bibitem{ielmini_-memory_2018}
D.~Ielmini and H.-S.~P. Wong, ``{In-memory computing with resistive switching
  devices},'' \emph{Nature Electronics}, vol.~1, no.~6, pp. 333--343, 2018.

\bibitem{sebastian_memory_2020}
A.~Sebastian, M.~Le~Gallo, R.~Khaddam-Aljameh, and E.~Eleftheriou, ``{Memory
  devices and applications for in-memory computing},'' \emph{Nature
  Nanotechnology}, vol.~15, no.~7, pp. 529--544, 2020.

\bibitem{burr_neuromorphic_2017}
G.~W. Burr, R.~M. Shelby, A.~Sebastian, S.~Kim, S.~Kim, S.~Sidler, K.~Virwani,
  M.~Ishii, P.~Narayanan, A.~Fumarola, L.~L. Sanches, I.~Boybat, M.~Le~Gallo,
  K.~Moon, J.~Woo, H.~Hwang, and Y.~Leblebici, ``{Neuromorphic computing using
  non-volatile memory},'' \emph{Advances in Physics: X}, vol.~2, no.~1, pp.
  89--124, 2017.

\bibitem{roy_towards_2019}
K.~Roy, A.~Jaiswal, and P.~Panda, ``{Towards spike-based machine intelligence
  with neuromorphic computing},'' \emph{Nature}, vol. 575, no. 7784, pp.
  607--617, 2019.

\bibitem{chi_prime_2016}
P.~Chi, S.~Li, C.~Xu, T.~Zhang, J.~Zhao, Y.~Liu, Y.~Wang, and Y.~Xie,
  ``{{PRIME}: A Novel Processing-in-Memory Architecture for Neural Network
  Computation in {ReRAM}-Based Main Memory},'' \emph{Int'l Symp. on Computer
  Architecture (ISCA)}, 2016.

\bibitem{geiger_larq_2020}
L.~Geiger and P.~Team, ``{Larq: An Open-Source Library for Training Binarized
  Neural Networks},'' \emph{Journal of Open Source Software}, vol.~5, no.~45,
  p. 1746, 2020.

\bibitem{krizhevsky_imagenet_2017}
A.~Krizhevsky, I.~Sutskever, and G.~E. Hinton, ``{{ImageNet} Classification
  with Deep Convolutional Neural Networks},'' \emph{Communications of the
  {ACM}}, vol.~60, no.~6, pp. 84--90, 2017.

\bibitem{he_deep_2016}
K.~He, X.~Zhang, S.~Ren, and J.~Sun, ``{Deep Residual Learning for Image
  Recognition},'' \emph{{IEEE} Conf. on Computer Vision and Pattern Recognition
  ({CVPR})}, 2016.

\bibitem{huang_densely_2017}
G.~Huang, Z.~Liu, L.~Van Der~Maaten, and K.~Q. Weinberger, ``{Densely Connected
  Convolutional Networks},'' \emph{{IEEE} Conf. on Computer Vision and Pattern
  Recognition ({CVPR})}, 2017.

\bibitem{guo2022join}
N.~Guo, J.~Bethge, C.~Meinel, and H.~Yang, ``{Join the High Accuracy Club on
  ImageNet with A Binary Neural Network Ticket},'' \emph{arXiv preprint
  arXiv:2211.12933}, 2022.

\bibitem{zhang2023binarized}
Y.~Zhang, A.~Garg, Y.~Cao, {\L}.~Lew, B.~Ghorbani, Z.~Zhang, and O.~Firat,
  ``{Binarized Neural Machine Translation},'' \emph{arXiv preprint
  arXiv:2302.04907}, 2023.

\bibitem{liu2020reactnet}
Z.~Liu, Z.~Shen, M.~Savvides, and K.-T. Cheng, ``{ReActNet: Towards Precise
  Binary Neural Network with Generalized Activation Functions},''
  \emph{European Conf. on Computer Vision ({ECCV})}, 2020.

\bibitem{Zhang_2022_CVPR}
Y.~Zhang, Z.~Zhang, and L.~Lew, ``{PokeBNN: A Binary Pursuit of Lightweight
  Accuracy},'' \emph{{IEEE} Conf. on Computer Vision and Pattern Recognition
  ({CVPR})}, 2022.

\end{thebibliography}
